\documentclass[aps,twocolumn,showpacs,superscriptaddress,preprintnumbers,nofootinbib,
amsmath,amssymb]{revtex4-1}

\usepackage{graphicx} 
\usepackage{tikz}
\usetikzlibrary{arrows}
\usepackage{multirow}
\usepackage[colorlinks=true,urlcolor=blue,linkcolor=blue,citecolor=blue]{hyperref}
\usepackage[normalem]{ulem}
\usepackage[capitalize]{cleveref}
\usepackage{amsfonts,amssymb}
 
\usepackage{comment}
\usepackage{flushend}

\usepackage[T1]{fontenc} 
\graphicspath{{fig/}}
\usepackage{mathtools}
\usepackage{xcolor}
\usepackage{graphicx}
\usepackage{subfigure}
\usepackage{dcolumn}
\usepackage{bm}

\definecolor{myblue}{RGB}{0, 100, 200}
\definecolor{myred}{RGB}{214, 39, 40}
\definecolor{mygreen}{RGB}{44, 160, 44}
\definecolor{mybrown}{RGB}{123, 64, 26}
\definecolor{mydarkblue}{RGB}{44, 77, 118}

\newcommand{\dd}{\mathrm{d}}

\begin{document}

\title{Dark Matter–Independent Orbital Decay Bounds on Ultralight Bosons from OJ287
}

\author{Qianhang Ding}
\email{dingqh@ibs.re.kr}
\affiliation{Cosmology, Gravity and Astroparticle Physics Group, Center for Theoretical Physics of the Universe, Institute for Basic Science (IBS), Daejeon, 34126, Korea}

\author{Minxi He}
\email{heminxi@ibs.re.kr}
\affiliation{Particle Theory and Cosmology Group, Center for Theoretical Physics of the Universe, Institute for Basic Science (IBS), Daejeon, 34126, Korea}

\author{Volodymyr Takhistov}
\email{vtakhist@post.kek.jp}
\affiliation{International Center for Quantum-field Measurement Systems for Studies of the Universe and Particles (QUP), High Energy Accelerator Research Organization (KEK), 1-1 Oho, Tsukuba, Ibaraki 305-0801, Japan}
\affiliation{Theory Center, Institute of Particle and Nuclear Studies (IPNS), High Energy Accelerator Research Organization (KEK), 1-1 Oho, Tsukuba, Ibaraki 305-0801, Japan}
\affiliation{Graduate University for Advanced Studies (SOKENDAI), 1-1 Oho, Tsukuba, Ibaraki 305-0801, Japan}
\affiliation{Kavli Institute for the Physics and Mathematics of the Universe (WPI), UTIAS, The University of Tokyo, Kashiwa, Chiba 277-8583, Japan}
 
\author{Hui-Yu Zhu}
\email{hzhuav@ibs.re.kr}
\affiliation{Cosmology, Gravity and Astroparticle Physics Group, Center for Theoretical Physics of the Universe, Institute for Basic Science (IBS), Daejeon, 34126, Korea}

\begin{abstract}
Ultralight bosons, predicted in scenarios beyond the Standard Model and viable dark matter (DM) candidates, can form superradiant clouds around spinning black holes influencing their dynamics.  
Using century-long monitored OJ287 supermassive black hole binary we set first DM–independent, dynamical constraints on their masses $\mu = (8.5-22) \times 10^{-22}$ eV. These dynamical constraints, driven by boson cloud friction, are robust against DM-model uncertainties and offer a novel ultralight boson probe. We show that analogous superradiant dynamics across the cosmic population of supermassive black hole systems could help resolve final-parsec evolution stalling problem and imprint a detectable suppression and break in the gravitational wave background. 
\end{abstract}

\maketitle

\section{Introduction}
Ultralight bosons (ULBs) with masses below the
eV-scale naturally arises in many extensions of the
Standard Model (SM) and constitute viable dark matter (DM) candidates (see, e.g.~\cite{Peccei:1977hh,Weinberg:1977ma,Wilczek:1977pj,Arvanitaki:2009fg,Turner:1983he,Press:1989id,Sin:1992bg,Hu:2000ke,Goodman:2000tg,Chikashige:1980ui,Gelmini:1980re}). 
With feeble ULB non-gravitational couplings, cosmological and astrophysical probes (see  e.g.~\cite{Kimball:2023vxk}) can be highly sensitive beyond and complementary to laboratory searches. 
When the wavelength of ULB is much larger than the black hole (BH) horizon, a rapidly rotating BH can efficiently amplify the bosonic field waves extracting rotational energy and
forming long-lived macroscopic ``gravitational atom'' (GA) clouds~\cite{Arvanitaki:2009fg,Arvanitaki:2010sy,Brito:2015oca,Baumann:2019eav}.
This superradiant amplification is the 
gravitational analogue of observed Dicke superradiance in quantum optics and condensed matter~\cite{Dicke:1954zz}. 
The resulting rich phenomena~\cite{yoshino2014gravitational,Chan:2022dkt,Ng:2020ruv,Stott:2020gjj,Davoudiasl:2019nlo,Baumann:2019ztm,Ding:2020bnl,Tomaselli:2023ysb,Li:2025qyu} offer intriguing windows for exploring fundamental physics.

In this work we introduce a novel dynamical probe of ULBs based on their superradiance impact on BH binary orbits, independent of any DM assumptions. Unlike previous ULB superradiance constraints derived from BH spin measurement statistics~(e.g.~\cite{Davoudiasl:2019nlo,Saha:2022hcd}), our method uses orbital decays as a new dynamical channel to probe the $\mu\sim10^{-21}$~eV window.  We further show that superradiant boson clouds can accelerate supermassive BH binary evolution in the late stage final-parsec regime, where evolution can problematically stall. An ideal system for this test is the OJ287 supermassive BH (SMBH) binary~\cite{Deb:2025raq}, monitored for over a century. Prior analyses of OJ287 crucially relied on new fields constituting DM abundance~\cite{Alachkar:2022qdt,2024ApJ...962L..40C,Deb:2025raq}.
Here, we set the first DM–independent dynamical limits on ULBs from SMBH orbital evolution, and identify distinctive gravitational wave background signatures, connecting particle physics, astrophysics, and gravitational wave (GW) cosmology.

\section{Ultralight boson superradiance}

ULB fields can efficiently seed macroscopic superradiant clouds around spinning BHs when their Compton wavelength $\lambda_c = 1/\mu$, with $\mu$ being their mass, exceeds Schwarzschild radius $r_s = 2 M_{\rm BH}$ of BH with mass $M_{\rm BH}$. As we show, superradiance clouds can significantly impact SMBH binary merger dynamics. We adopt $c=\hbar=G=1$.

Let us consider a minimal scenario of massive scalar field $\Phi$ propagating on a 
Kerr spacetime background. with Lagrangian contributions $\mathcal{L} \supset   
-\left( \partial \Phi \right)^2/2 -\mu^2 \Phi^2/2 $, neglecting possible self-interactions (see~\cite{Arvanitaki:2010sy,Baryakhtar:2020gao,Omiya:2022gwu,Chia:2022udn,Xie:2025npy,Lambiase:2025twn} for their potential effects).
The resulting Klein-Gordon equation governing the ULB behavior 
can be simplified to a Schr\"{o}dinger-like equation by expanding it with respect to a small ``gravitational fine structure constant'' $ \alpha = M_{\rm BH} \mu < 1 $. See App.~\ref{Sec.GA_Detail} for details. 
In this regime, the Schwarzschild radius of BH is smaller than the Compton wavelength of ULB and the system can be described as ``gravitational atom (GA)''
resembling eigenfunctions of a hydrogen atom described by quantum numbers $n$, $l$, $m$ as
\begin{align}
\psi_{nlm}(r,\theta,\phi)\simeq R_{nl}(r)Y_{lm}(\theta,\phi)e^{-i(\omega_{nlm}-\mu)t} ~,
\end{align}
where $R_{nl}(r)$ is the radial wave function and $Y_{lm}(\theta,\phi)$ are the spherical harmonics. 
The corresponding eigenvalue consists of a real part and an imaginary part $\omega_{nlm}=E_{nlm}+i\Gamma_{nlm}$, which denotes the energy of the eigenstate and its growth or decay rate respectively.  
Exponential superradiance growth occurs when $ \Gamma_{nlm} >0$.

The superradiant growth rate 
of an $|nlm\rangle$ state
can be computed analytically using the Detweiler approximation~\cite{Detweiler:1980uk}, considering $ \alpha \lesssim 0.3 $~\cite{Brito2015b,Cannizzaro:2023jle},
\begin{align}
    \Gamma_{nlm}=2\, \tilde{r}_+\, C_{nl} \, g_{lm}(m\, \Omega_{\rm BH}-\omega_{nlm})\, \alpha^{4l+5}~,
\end{align}
where $\Omega_{\rm BH}=\chi/2M_{\rm BH}\tilde{r}_+$ is the angular velocity of the outer BH horizon at $\tilde{r}_+=1+\sqrt{1-\chi^2}$ and 
the dimensionless BH Kerr spin parameter is defined as $\chi = J/M_{\rm BH}^2$, where $J$ being the angular momentum.
Here, $C_{nl}$ are numerical coefficients and  $g_{lm} = g_{lm}(\chi,\alpha,\mu)$ are functions~\cite{Detweiler:1980uk,Baumann:2019eav}. 
In a superradiant state with $\Gamma_{nlm} > 0$,  
ULB cloud will exponentially grow until spin $ \chi$ becomes insufficient and saturates at 
\begin{align}
    \chi_{\text{sat}}=\frac{4m\alpha}{m^2+4\alpha^2}~.   
\end{align}

\begin{figure}[t] 
\centering
   \includegraphics[width=\linewidth]{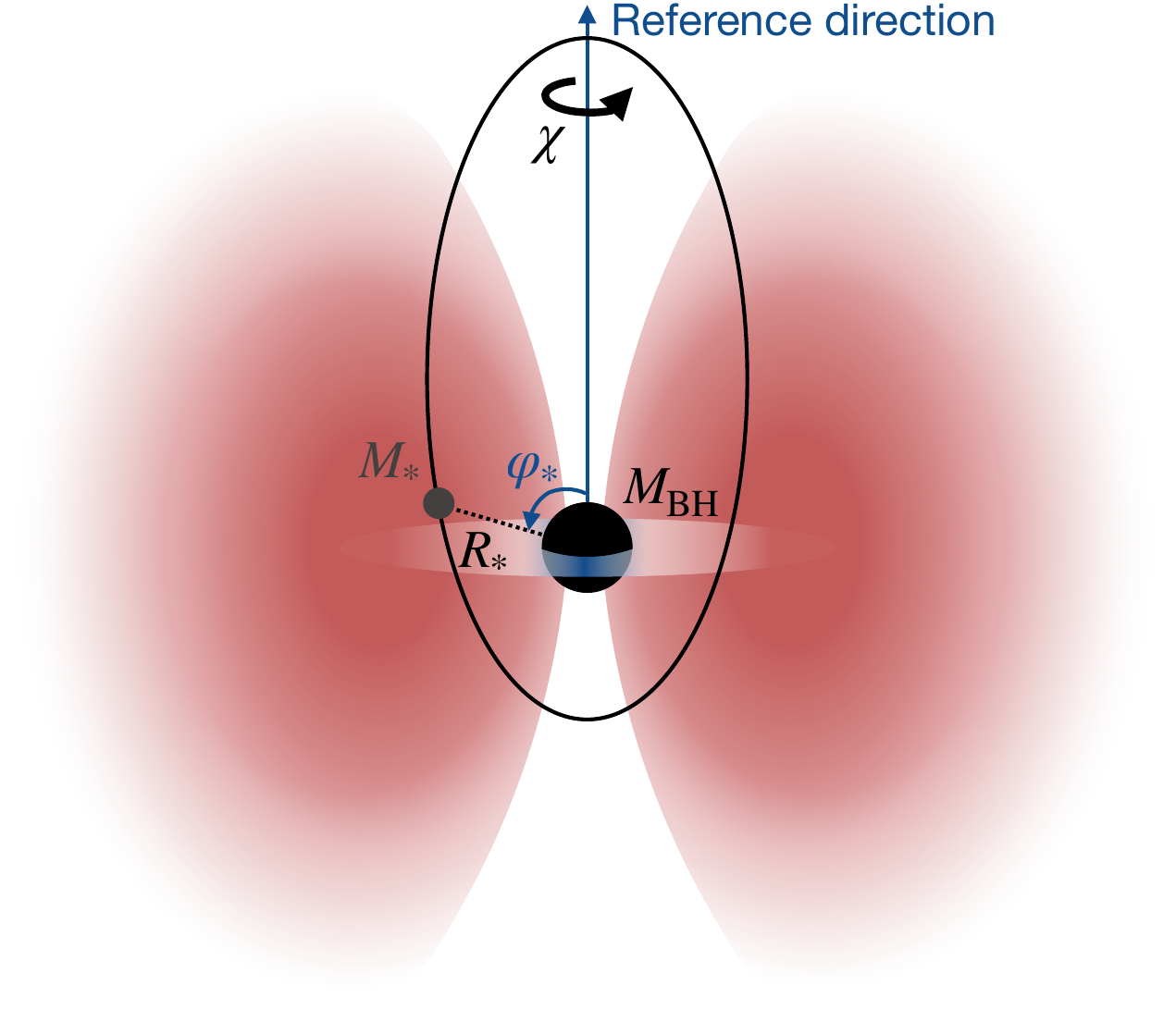}
	\caption{Schematic illustration of secondary BH (gray) of mass $M_{\ast}$ (black) traversing ULB superradiance cloud in $|211\rangle$ state orbiting primary BH of mass $M_{\rm BH}$ and spin $\chi$ in SMBH binary, such as OJ287 system. Primary BH rotation defines the reference axis. The secondary’s position is given by its separation $R_{\ast}$ and orbital phase $\varphi_{\ast}$ measured from this axis. The orbital angular momentum is perpendicular to the primary BH spin. 
	}
    \label{fig:illustration}
\end{figure}

\begin{figure}[t]
\centering
\includegraphics[width=\linewidth]{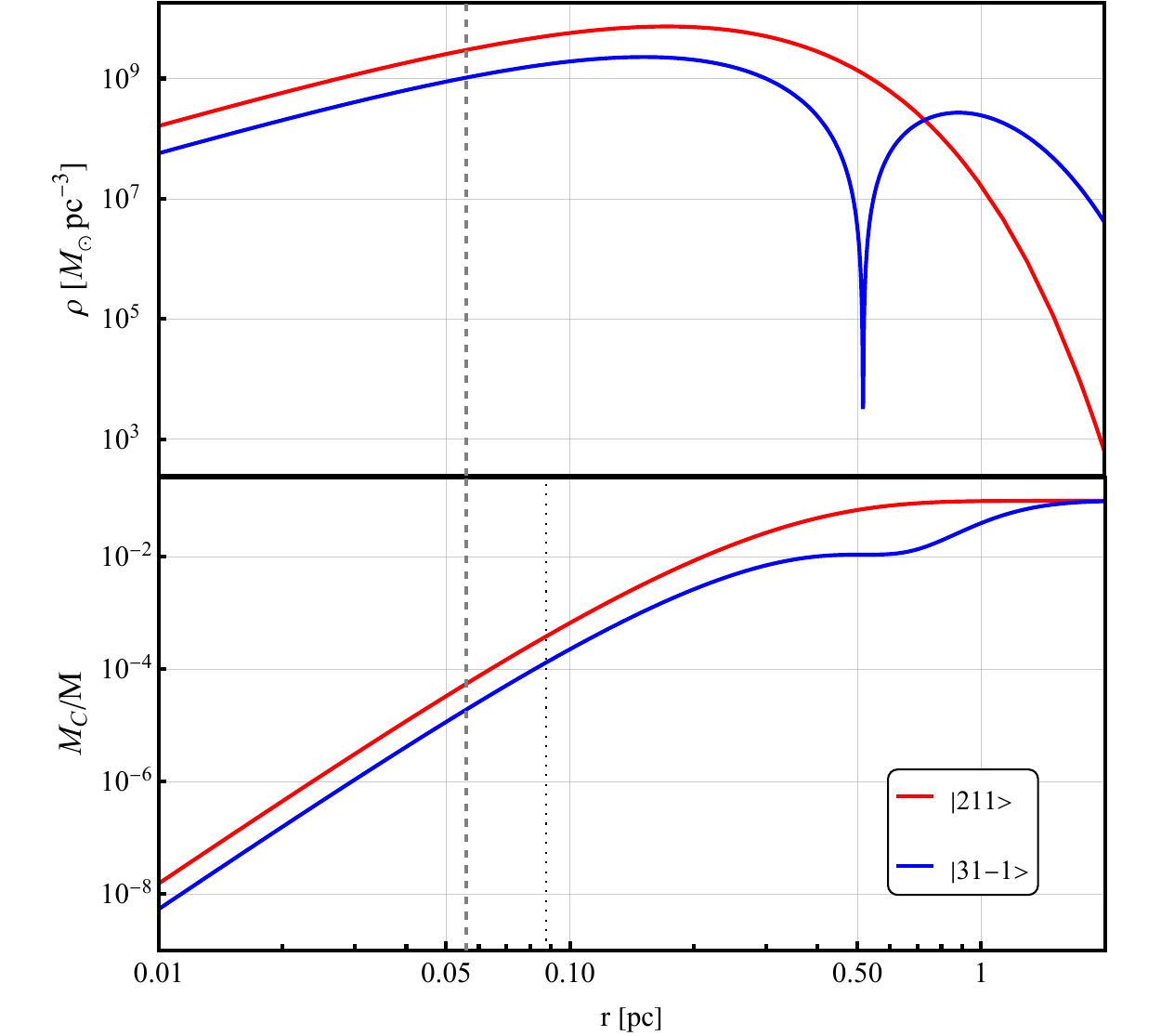}
\caption{[Top] Density profiles of the $|211\rangle$
and $|31-1\rangle$ GA states for $\alpha=\beta=0.1$ around primary BH in OJ287 system.  
[Bottom] Ratio of enclosed superradiance cloud mass relative to SMBH $M_{\rm BH}$.
The vertical dashed line mark the semi-major axis
($0.056$\,pc) of OJ287. The dotted line corresponds to $50\,r_s$.}
\label{fig:DensityProf}
\end{figure}

By analogy with electric case, GA Bohr radius is $r_0 = 1/\mu \alpha$, and the excited states have radii $r_n \simeq n^2 r_0$. The density of a boson cloud $\rho_{nlm}= M_{nlm} |\psi_{nlm}|^{2}$  can be described as (e.g.~\cite{Cao:2023fyv})  
\begin{align}\label{eq:density_profile} 
    \rho_{nlm}(x, \theta) =&~  2.72 \times 10^{12} \, \mathrm{M_\odot \,pc^{-3}}  A_{nlm} (x, \theta) \frac{\beta}{0.1} \left(\frac{\alpha}{0.1}\right)^6 \notag\\ &\times   \left(\frac{1.83\times10^{10}M_\odot}{M_{\rm BH}}\right)^2~,   
\end{align}
where $A_{nlm}(x, \theta)$ denotes the spatial density distribution, where $x = r/r_0$ 
and $\beta$ is the mass ratio between the boson cloud and BH. 
For the dominant states in this work
\begin{align}\label{eq:211_density}
A_{211}(x,\theta)&= \frac{1}{64\pi}   x^{2}e^{-x} \sin^{2}\theta~, \\
A_{31-1}(x,\theta)&=\frac{1}{6561\pi} (x^{2}-6x)^{2}e^{-2x/3} 
\sin^{2}\theta~.
\end{align}
In the following discussion, we consider the impact of superradiant bosons in OJ287 orbital dynamics as illustration in Fig.~\ref{fig:illustration}. 
The density profile of superradiant bosons plays an important role in determining the orbital dynamics of OJ287, we display the corresponding profiles and
enclosed cloud mass for parameters relevant to OJ287 
($\alpha=\beta=0.1$) in Fig.~\ref{fig:DensityProf}. 

\section{OJ287 dynamics and novel boson limits}
A pristine SMBH system to explore the dynamical effects of superradiance clouds is blazar OJ287.  
OJ287 is an unusually massive and tight BH binary with primary BH mass of $M_{\rm BH} = 1.83\times10^{10} \, M_\odot$, secondary BH mass of $M_{\ast} = 1.50\times 10^{8} \, M_\odot$ and orbital timing measurements collected over $\sim 120$\,yr period. 
Its orbital parameters include semi-major axis $a = 0.056 \, \mathrm{pc}$, inclination angle $\iota=90^\circ$, and eccentricity $e = 0.657$~\cite{Dey:2018mjg, 1997ApJ...481L...5V,Valtonen:2013spa} as illustration in 
The binary is shrinking due to continuous GW emission, expected from general relativity~\cite{Peters:1964zz}
\begin{equation}
\langle P_{\rm GW} \rangle = -\dfrac{32 m_r^2 M_t^3}{5 a^5}(1-e^2)^{-7/2}\Big(1 + \frac{73}{24}e^2 + \dfrac{37}{96}e^4\Big)~,
\end{equation}
where $m_r = M_{\rm BH} M_{\ast}/(M_{\ast} + M_{\rm BH})$ is the reduced system mass and $M_t = M_{\ast} + M_{\rm BH}$ is the total mass.

Recent work claimed an apparent OJ287 total orbital-decay power  
$\langle P_{\rm tot} \rangle= (3.66\pm0.24)\times10^{41}\,$W seemingly exceeds the quadrupole GW prediction  
$\langle P_{\rm GW} \rangle=(2.62\pm0.02)\times10^{41}\,$W by $\sim 4.3\sigma$~\cite{2024ApJ...962L..40C}.
However, detailed reanalysis finds only a $0.41\sigma$ deviation~\cite{Deb:2025raq}.  
Remaining conservative, we base our limits on the null-excess result compared to relativity GW predictions and derive the first DM-independent, dynamical constraint on ULBs from OJ287.

ULBs of mass $ \mathcal{O}(10^{-21}) \,\mathrm{eV}$ can form a superradiance cloud around primary SMBH of OJ287 within the current age of the Universe  $\tau_\mathrm{U} \simeq 13.8 \, \mathrm{Gyr}$, considering high initial spin $\chi_i$ as expected of most SMBHs~\cite{Volonteri:2004cf}.
This can be seen from the superradiance growth rate of the fastest growing GA state $|211\rangle$ 
\begin{equation}
    \Gamma_{211}  \simeq  1.2 
    \times 10^{-16}  \left( \frac{\alpha}{0.1} \right)^9  \left( \frac{1.8 \times 10^{10} M_\odot}{M_{\rm BH}} \right) \, {\rm s}^{-1}~ .
\end{equation}
The $|211\rangle$ cloud has a typical Bohr radius of $ r_0 \simeq \mathcal{O}(0.01-1) \, \mathrm{pc}$ and density of $ \rho_{211} $. 
Hence, $|211\rangle$ GA state can be fully saturated and with the cloud mass of $M_{211} \simeq \alpha M_{\rm BH}$~\cite{Tsukada:2018mbp}.
The total cloud growth time takes number of e-folds that can be up to
$\log (\Delta \chi M_\text{BH}^2) \sim 170$ times larger than $1/\Gamma_{211}$, which is still significantly less than $\tau_U$ for OJ287.

The superradiant cloud could potentially deplete through variety of channels, such as monochromatic GW emission~\cite{Brito:2014wla,Yoshino:2013ofa} or Landau-Zener bound-to-bound resonances~\cite{Baumann:2019ztm,Ding:2020bnl,Takahashi:2021yhy,Tong:2021whq,Takahashi:2023flk} and bound-to-unbound resonance, known as ionization~\cite{Baumann:2021fkf,Tomaselli:2023ysb}. Detailed analysis demonstrates (see App.~\ref{Sec.Binary_Companion} for details) that the OJ287 system remains not significantly affected even when these effects are taken into account.

As the secondary BH of mass $M_{\ast}$ traverses superradiance GA cloud of local density $\rho$, experiences a gravitational drag force acting as dynamical friction~\cite{1943ApJ....97..255C}, which in the non‐relativistic, wave‐like regime reads~\cite{Hui:2016ltb}
\begin{equation}
  F_{\rm DF}
  = \frac{4\pi\,M_{\ast}^2\,\rho}{v^2}\,C_\Lambda(\xi,k\,r_\Lambda)\,,
  \label{eq:DF_force}
\end{equation}
where $v$ is the speed relative to the boson cloud that equals $\alpha \sqrt{(1+q)/x_{\ast}}$ in circular orbit and $\alpha \sqrt{(1+q)(2/x_{\ast}-r_0/a)}$ in elliptic orbit with $x_{\ast}= R_{\ast}/r_0$, $\xi= M_\ast\mu/v$, $k=\mu v$, and $r_\Lambda$ representing the smaller quantity between the size of the orbit and the size of the cloud. In the $\xi \ll 1$ limit the Coulomb logarithm becomes
\begin{align}
    C_\Lambda(kr) = \mathrm{Cin}(2kr) + \frac{\sin 2kr}{2kr} - 1~,
\end{align}
with
\begin{align}
    \mathrm{Cin}(z) = \int_0^z (1 - \cos t) \frac{\dd t}{t}~.
\end{align}

Writing the instantaneous frictional power as $P_{\rm DF}=F_{\rm DF}\,v$, and $q=M_{\ast}/M_{\rm BH}$ being the binary mass ratio, one finds
\begin{equation}
  P_{\rm DF}
  = 4\pi\,\frac{q^2 M_{\rm BH}^2\,\rho (x_{\ast},\theta_{\ast})}{ \alpha\sqrt{(1+q)(2/x_{\ast}-r_0/a)}}  \,
    C_\Lambda\!\bigl(k\,r_\Lambda\bigr).
  \label{eq:DF_power}
\end{equation}
For eccentric orbits with semi-major axis $a$, eccentricity $e$, and eccentric anomaly $z$ with $r=a[1-e\cos z]$, the effective impact parameter $kr_\Lambda$ is given by \cite{Zhang:2019eid}
\begin{equation}
  kr_\Lambda = 
  \begin{cases}
    \sqrt{x_p}\,\dfrac{(1-e\cos z)^{3/2}}{\sqrt{1+e\cos z}}\,, 
      & x_p\le x_{97},\\[6pt]
    \dfrac{x_{97}}{\sqrt{x_p}}\,\dfrac{(1-e\cos z)^{3/2}}{\sqrt{1+e\cos z}}\,, 
      & x_p>x_{97},
  \end{cases}
\end{equation}
where $x_p=r_p/r_0$ is the periapsis in units of $r_0$, and $x_{97}$ encloses 97\% of the cloud mass.

To estimate the impact of boson cloud-induced dynamical friction on SMBH binary evolution, we compute dynamical friction power averaged over one period. The orbit of SMBH binary can be described as
\begin{equation}
    R_{\ast}(\varphi_{\ast}) = \frac{a(1-e^2)}{1+e \cos{(1-\zeta)\varphi_{\ast}}}~,
\end{equation}
where   $\zeta$ is the precession phase angle per period.
When a secondary SMBH companion is traversing the GA boson cloud, it experiences position-dependent cloud density $\rho(R_{\ast},\theta_{\ast})$. The angular position $\theta_{\ast}$ depends on the orbital phase $\varphi_{\ast}$ and inclination angle $\iota$ as
$   \cos{\theta_{\ast}} = \cos{\varphi_{\ast}} \sin{\iota}$.
The detailed OJ287 parameters are listed in Table.~\ref{Table.Para_app}.
\begin{table}[t]
\centering
\begin{tabular}{    c | c | c    } 

 \hline\hline
 Parameters & Value & Ref. \\
 \hline
 Primary SMBH mass $(10^{10}M_{\odot})$   &  $1.8348 \pm 0.0008$ & \cite{Dey:2018mjg} \\
 Secondary SMBH mass $(10^8M_{\odot})$ &  $1.5013 \pm 0.0025$ & \cite{Dey:2018mjg}\\
 Eccentricity $e$ & $0.657 \pm 0.001$ & \cite{Dey:2018mjg}\\
 Precession angle per period $\Delta \vartheta$   & $38.62^\circ \pm 0.01^\circ$ & \cite{Dey:2018mjg}\\
 Orbital period $T$ (yr) & $12.067 \pm 0.007$ & \cite{Dey:2018mjg}\\
 Orbital period decay rate $\Dot{T}$ & $-0.00099 \pm 0.00006$ & \cite{Dey:2018mjg}\\
 Spin of the primary SMBH $\chi_{P_2}$ & $0.381 \pm 0.004$ & \cite{Dey:2018mjg}\\
 Spin of the primary SMBH $\chi_{P_1}$ & $0.313 \pm 0.01$ & \cite{Valtonen:2016awd}\\
 Semi-major axis $a$ (pc) & $0.056$ & \cite{1997ApJ...481L...5V}\\
 Inclination angle of orbit plane & $90^{\circ}$ & \cite{Valtonen:2013spa}\\ 
 \hline \hline
\end{tabular}
\caption{Parameters characterizing OJ287 BH system.}
\label{Table.Para_app}
\end{table}

Then, the average of Eq.~\eqref{eq:DF_power} over one orbital period $T$ is
\begin{equation}\label{eq:average_DF_power_app}
    \langle P_{\rm DF}\rangle
    = \frac{1}{T}\int_0^{2\pi} P_{\rm DF}(R_{\ast}(\varphi_{\ast}), \theta_{\ast}(\varphi_{\ast},\iota))\,
    \frac{\dd t}{\dd\varphi_{\ast}}\,\dd\varphi_{\ast}~.
\end{equation}
Here,
\begin{equation}
    \frac{\dd \varphi_{\ast}}{\dd t} = \frac{\sqrt{(1+q)M_{\rm BH} a(1-e^2)}}{R_{\ast}(\varphi_{\ast})^2}~
\end{equation}
and the orbital period $T$ can be calculated as
\begin{equation}
    T = \int_0^{2\pi} \frac{\dd t}{\dd \varphi_{\ast}} \dd \varphi_{\ast}~.
\end{equation}
We employ the resulting Eq.~\eqref{eq:average_DF_power_app} to compute the orbital decay rate due to ULB clouds.
\begin{figure}[t] \centering
   \includegraphics[width=\linewidth]{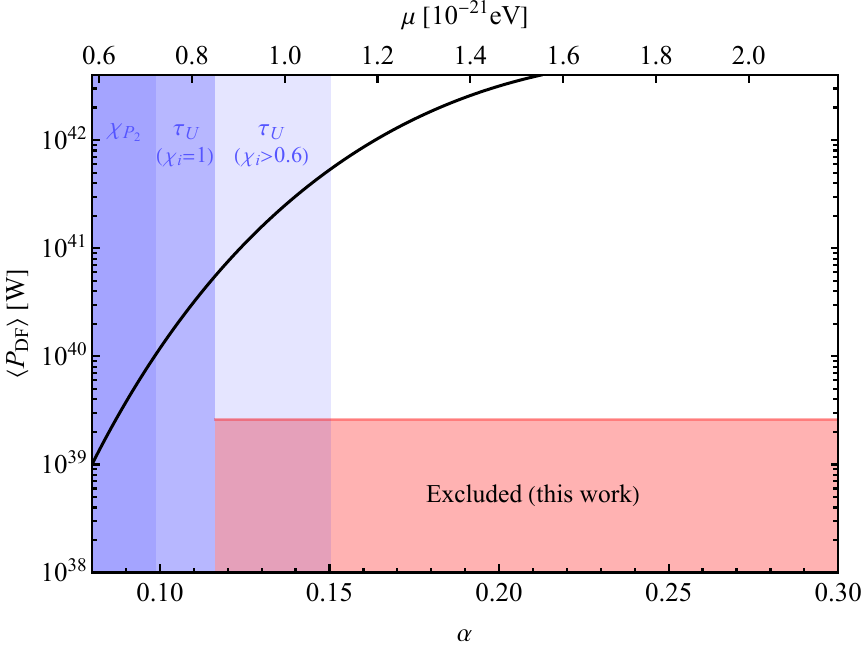}
	\caption{  
Orbital decay power from dominant superradiance cloud state in OJ287 and novel upper limits considering null excess over relativity GW predictions $\langle P_{\rm DF} \rangle/\langle P_{\rm GW} \rangle < 0.01$ (red region) from dynamical friction. Region where saturated spin is below the measured primary BH spin value $\chi_{P_{2}}$ (light blue region) is shown. Parameters where cloud won't form during cosmic time $\tau_{\rm U}$ (pink region), considering wide initial spin range $\chi_i$, are shown.  
	}
    \label{fig:DF_power_OJ287}
\end{figure}
In Fig.~\ref{fig:DF_power_OJ287} we display computed orbital-decay power generated by dynamical friction in OJ287 due to a saturated GA $|211\rangle$ state. While our baseline analysis assumes no excess beyond emission predicted by relativity, our framework can also accommodate a genuine deviation if such is confirmed. For GA couplings $\alpha=0.121 - 0.127$, corresponding to boson masses
$\mu=(8.8-9.3)\times10^{-22}~$eV, the predicted dissipation can accommodate claimed excess in orbital decay power $\langle P_{\rm DF} \rangle = \langle P_{\rm tot} \rangle - \langle P_{\rm GW} \rangle$ of~\cite{2024ApJ...962L..40C}.  In our analysis we conservatively adopt null–excess~\cite{Deb:2025raq}, and derive new DM–independent bounds. 
Here, the superradiance growth time obeys $170 \times\Gamma_{211}^{-1}<  \tau_\mathrm{U}$ and the primary spin satisfies $\chi<\chi_{211,{\rm sat}}$, with $\chi_{P_{2}}$ being more conservative compared to considering $\chi_{P_{1}}$. Hence, the $|211\rangle$ state is ensured to be populated and stable. See App.~\ref{Sec.Binary_Companion} for further details.
  
Throughout this band the superradiance growth time obeys $170 \times\Gamma_{211}^{-1}<  \tau_\mathrm{U}$ and the primary spin satisfies $\chi<\chi_{211,{\rm sat}}$, with $\chi_{P_{2}}$ being more conservative compared to considering $\chi_{P_{1}}$. Hence, the $|211\rangle$ state is ensured to be populated and stable. See App.~\ref{Sec.Binary_Companion} for further details.

Our scenario is consistent with timing observations of OJ287 flares, which are thought to originate from secondary BH interacting with the primary BH accretion disk.  Integrating the density profile for the $|211\rangle$ state out to $50\,r_s$ yields a cloud mass of $\sim 0.05\% $  of the primary BH mass as shown in bottom panel of Fig.~\ref{fig:DensityProf}. This is  well below the $\sim 3\%$ mass ratio at which the flare timing predictions could be impacted, suggested from ~\cite{Alachkar:2022qdt}. 

We set novel limits on ULBs independent of assumptions on their DM modeling and cosmological abundance considering conservatively no excess of orbital-decay power over relativity GW predictions at $0.41\sigma$ at 1PN level~\cite{Deb:2025raq},
which corresponds with 
observational limit power ratio $\langle P_{\rm DF}\rangle/\langle P_{\rm GW} \rangle \lesssim 0.01$ in Fig.~\ref{fig:DF_power_OJ287}. Hence, this forbids ULB masses larger than $6.4\times10^{-22}\,\text{eV}$ considering superradiance effects. Combining with requirements that superradiant boson cloud efficiently forms satisfying $170 \times\Gamma_{211}^{-1}<  \tau_\mathrm{U}$, $\chi<\chi_{211,{\rm sat}}$, and Detweiler approximation validity range $\alpha \lesssim 0.3$, we set novel bounds restricting ULB mass range $\mu \simeq (8.5-22)\times 10^{-22}$~eV. Since additional model‑dependent effects such as ULB self‑interactions only suppress superradiant growth, our derived limits are conservative. Varying OJ287 parameters within 1$\sigma$ uncertainties that we consider   excluded ULB limit region is not significantly affected. 

Our ULB constraints are the first derived from OJ287 orbital dynamics that are independent of DM-assumptions, unlike previous studies~\cite{Alachkar:2022qdt,2024ApJ...962L..40C,Deb:2025raq}.

\section{Final Parsec Evolution}\label{App:final_parsec}
GA superradiance clouds stemming from ULBs of mass around $\mu\simeq \mathcal{O}(10^{-21})\,\text{eV}$, as suggested by OJ287 orbital decay analysis, envelop
$\mathcal O(10^{10})\,M_\odot$ SMBHs with Bohr radii $r_c\sim\mathcal{O}(1)$ pc. At this regime the evolution of 
SMBH binaries could stall, known as the ``final-parsec problem'', denoting inefficient SMBH binary passage through its final parsec separation within a Hubble timescale~\cite{Milosavljevic:2002ht}. Recently reported pulsar-timing arrays (PTAs) detection of a stochastic GW background in the nanohertz band~\cite{NANOGrav:2023gor,EPTA:2023fyk,Zic:2023gta,Xu:2023wog} that can originate from SMBHs further highlights these effects~\cite{InternationalPulsarTimingArray:2023mzf}.
Hence, the dynamical friction from GA boson clouds in SMBH binaries could efficiently accelerate  the shrinkage of binary separation within final parsec evolution.

\begin{figure}[t] \centering
        \includegraphics[width=8cm]{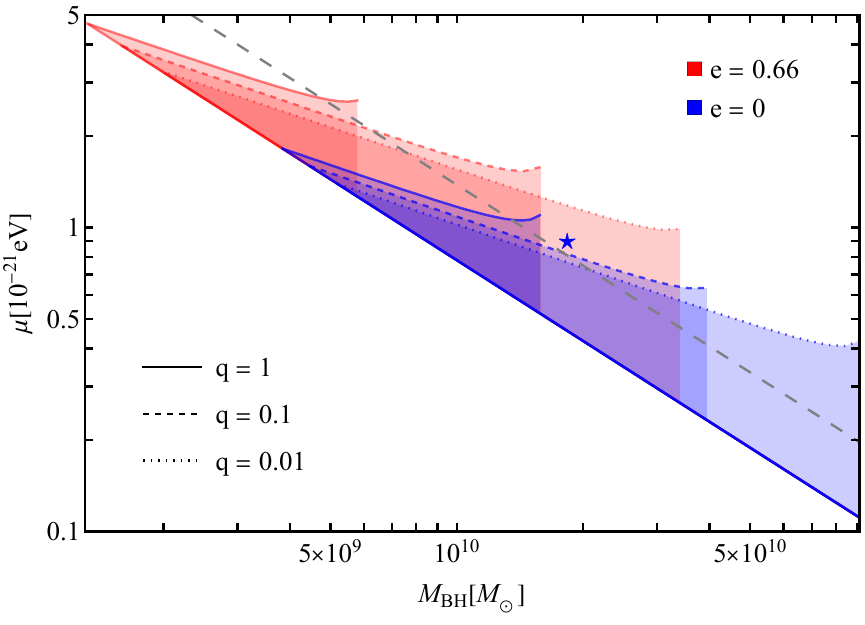}
        \includegraphics[width=8cm]{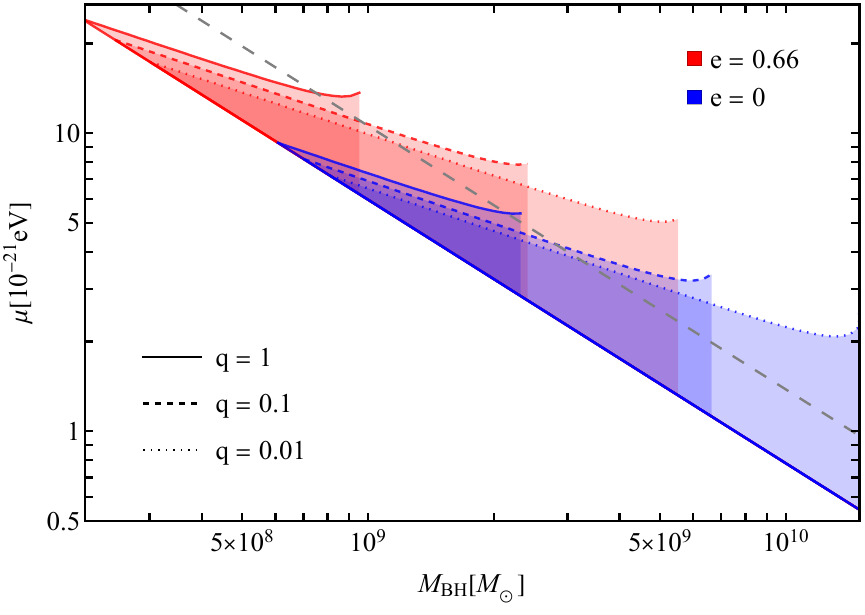}
	\caption{
    [Top] Allowed SMBH mass and ULB mass parameter space where binary evolution time from $2 \, \mathrm{pc}$ to $0.01 \, \mathrm{pc}$ is shorter than $ \tau_{\rm pc} < 1 \, \mathrm{Gyr}$, alleviating the final-parsec problem. The gray dashed line corresponds to conservative superradiance timescales of $170 \times \Gamma_{211}^{-1} < \tau_U$. The parameters for addressing OJ287 orbital decay $(1.83 \times 10^{10} \, M_\odot,\, 0.9 \times 10^{-21} \, \mathrm{eV})$ are shown by a blue star.
    [Bottom] Allowed SMBH mass and ULB mass parameter space where binary evolution time from $0.5 \, \mathrm{pc}$ to $0.1 \, \mathrm{pc}$ is shorter than $ \tau_{\rm pc} < 1 \, \mathrm{Gyr}$, alleviating the final-parsec problem.
	}
        \label{fig:Parameter}
\end{figure}

We estimate that bound-to-bound  resonance transitions deplete at most
$\mathcal{O}(10)\%$   of the $|211\rangle$ population for equal-mass binaries and  
$\lesssim10\%$ for $q=0.1$. See Fig.~\ref{fig:SurvRate} for further details. This is distinct from OJ287 where GA is taken to be almost purely in $|211\rangle$ state.
To capture this behavior we model the GA cloud distribution as  
$\rho=\eta\rho_{211}+(1-\eta)\rho_{31{-}1}$, with $\eta$ denoting relative contribution of $|211\rangle$ and $|31-1\rangle$ states that depends on SMBH binary parameters. For $q = 0.1$ binaries we have $\eta \simeq0.9$, while for $q = 1$ case $\eta \simeq 0.1$. Here, we estimate negligible contributions of other states, although our approach can be readily extended to incorporate this. 
Then the orbital decay rate of SMBH binaries can be found by semi-major axis evolution as
\begin{align}\label{eq:orbit_decay}
    \frac{1}{a}\frac{\dd a}{\dd t} = \frac{2 a}{q M_{\rm BH}^2} (\left<P_\mathrm{DF}\right> + \left<P_\mathrm{GW}\right>)~.
\end{align}
The resulting elapsed time within final parsec $\tau_{\rm pc}$ can be estimated by integrating Eq.~\eqref{eq:orbit_decay} over evolving separations to obtain its evolution time as follows,
\begin{equation}
    \tau = \int \frac{q M_{\rm BH}^2}{\langle P_\mathrm{DF} \rangle + \langle P_\mathrm{GW}\rangle} \frac{1}{2 a^2} \dd a~.
\end{equation}
Solving final-parsec problem requires the evolution time $\tau$ smaller than $1 \, \text{Gyr}$, and this put constraints on SMBH mass and boson mass. We consider the evolving separation from $2\,\text{pc}$ to $0.01\, \text{pc}$ in the top panel of Fig.~\ref{fig:Parameter}. Here, we partially release this evolving separation from $0.5\,\text{pc}$ to $0.1\, \text{pc}$, and the resulting allowed parameter regions are shown in the bottom panel of Fig.~\ref{fig:Parameter}. It shows the existence of superradiant boson cloud can help $\mathcal{O}(10^8 - 10^{10}) \, M_\odot$ SMBH binaries overcome their final $\mathcal{O}(0.1)\,\text{pc}$ separations.

In Fig.~\ref{fig:Parameter} we display the region in the
$(M_{\rm BH},\mu)$ plane for which the inspiral during final-parsec evolution is completed in
$\tau_{\rm pc}<1$ Gyr. Here, the lower curve denotes $\Gamma_{211}^{-1}<\tau_\mathrm{U}$, and with dashed line showing conservative $170 \times \Gamma_{211}^{-1}<\tau_\mathrm{U}$.
Since the parameters relevant for OJ287  
$\bigl(1.8\times10^{10}M_\odot, 0.9\times10^{-21}\,\text{eV}\bigr)$ lie inside the allowed band this indicates that
dynamical friction from the ULB cloud can drive the binary through the final parsec
before GW back-reaction takes over, and can simultaneously account for the claimed OJ287 orbital decay excess if confirmed. Here, we also display for reference the circular binary case with eccentricity $e = 0$.  Systems with ULB clouds that satisfy the superradiance constraint but reside outside of the shaded region will also partially accelerate the decay. This indicates that our scenario can alleviate the final-parsec problem in various SMBH systems.

\section{Gravitational wave background and turnover} 
Once a SMBH binary evolution has crossed the final parsec, GW emission dominates and sources
nanohertz stochastic GW background observable by PTAs. A superradiant cloud drag reduces the energy radiated
per logarithmic frequency interval,
\begin{align} \label{eq:dedrag}
    \frac{\dd E_\mathrm{GW}}{\dd f_s} = \frac{\pi^{2/3}}{3} \frac{M_{\rm BH}^{5/3}}{f_s^{1/3}} \frac{q}{(1+q)^{1/3}} \frac{\langle P_\mathrm{GW} \rangle}{\langle P_\mathrm{GW} \rangle + \langle P_\mathrm{DF} \rangle}~,
\end{align}
with a characteristic strain 
\begin{align}\label{eq:SGWB_strain}
    h_c^2(f) = \frac{4 }{\pi f} \int \dd z \dd q\dd M_{\rm BH}  \frac{\dd^3 n}{\dd z \dd q \dd M_{\rm BH}} \frac{\dd E_\mathrm{GW}}{\dd f_s}~,
\end{align}
where $f_s = f(1+z)$ is the source frequency and $\dd^3n/\dd z \dd q\dd M_{\rm BH}$ is the SMBH binary population that we estimate considering galaxy merger rate distribution, where we follow \cite{NANOGrav:2023hfp, Alonso-Alvarez:2024gdz} to evaluate SMBH population (see App.~\ref{app:SMBH_population}). 
In evaluating Eq.~\eqref{eq:dedrag} we account for distinct GA configurations by appropriately considering cloud distribution profile $\rho$ with ratio of states $\eta$ set depending on SMBH binary and ULB parameters.
With superradiant growth relevant when $\Gamma_{211}^{-1}<\tau_\mathrm{U}$, we approximate that only SMBHs above a critical mass $M_{\rm crit}$ host GA clouds for a given $\mu$.
Hence, dynamical friction acts for $M_{\rm BH}\gtrsim M_{\rm crit}$ in the mass integral range
$10^{5}-10^{10}M_\odot$. We denote with band parameter range extending to conservative $170\times\Gamma_{211}^{-1}<\tau_\mathrm{U}$. This includes contributions from partially populated boson clouds. Since significant fraction of SMBHs is expect to have large spin~\cite{Volonteri:2004cf}, we consider a scenario where all SMBHs are initial maximal spin, although our results can be readily linearly rescaled for smaller fraction.

\begin{figure}[t] \centering
        \includegraphics[width=\linewidth]{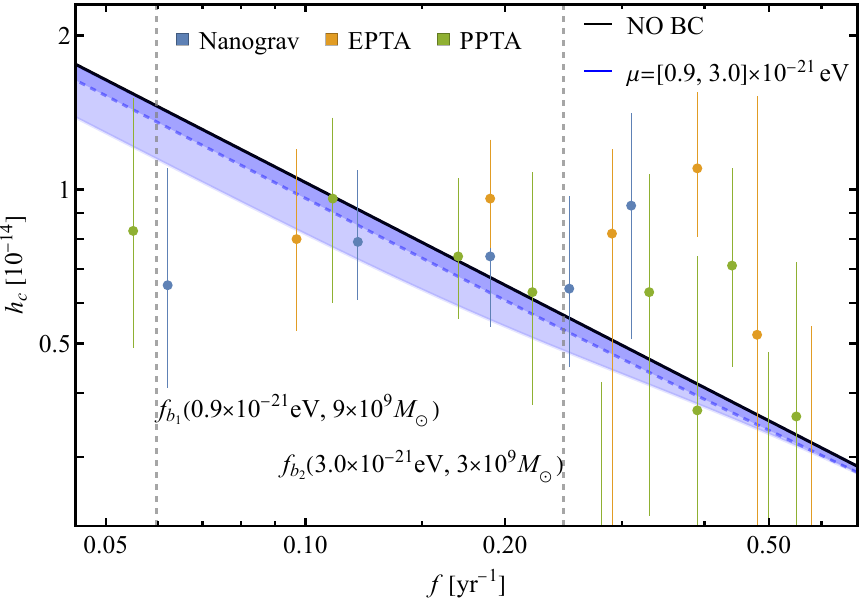}
	\caption{
GW background strain resulting from cosmic population of SMBH binaries considering ULB cloud dynamical friction effects for a range of boson masses. Black lines denotes standard relativity GW predictions. Shaded region above the blue solid line depicts superradiance effects with all SMBHs having maximal initial spin, and shaded region above blue dashed line depicts conservative case of $170 \times \Gamma_{211}^{-1} < \tau_U$. The vertical lines show new turnover GW frequency signatures considering $f_b$ for different $(\mu, M_{\rm BH})$ cases. Observations from NANOGrav~\cite{NANOGrav:2023gor}, EPTA~\cite{EPTA:2023xxk}, and PPTA~\cite{Reardon:2023gzh} are shown. }
        \label{fig:SGWB}
\end{figure}

In Fig.~\ref{fig:SGWB} we show that the GA cloud effects of our scenario can suppress the PTA strain by
$\mathcal{O}(10-30)\%$ relative to the vacuum prediction, improving agreement with the observed PTA amplitude, where the predicted strain in GA cloud scenario gives $\chi^2 = 16.5$ (blue solid  - all SMBHs with maximal spin), $\chi^2 = 17.5$ (blue dashed  - conservative superradiance timescales), while predicted strain in GW scenario gives $\chi^2 = 18.8$ (black solid -  general relativity prediction).(see App.~\ref{app:chi_squared} for statistical analysis).
For completeness we note that saturated GA clouds continuously radiate in GWs~\cite{Brito:2017wnc}. This lies in $\sim 0.1$ microhertz band for $10^{-21}\,\text{eV}$ ULBs and contributes negligibly to
the nanohertz background. 

Since the superradiant cloud drag becomes ineffective once the SMBH orbit shrinks inside
its Bohr radius $r_0$,
GW background exhibits a turnover at
\begin{equation}
f_b \simeq 3.2~ \mathrm{nHz}~\Big( \frac{\mu}{10^{-21}\,\mathrm{eV}} \Big)^3 \Big( \frac{M_{\rm BH}}{10^{10}\,M_\odot} \Big)^2~.
\end{equation}
This is a distinct testable feature of our scenario, distinguishing it from other proposals such as those based on DM spikes~(e.g.~\cite{2024ApJ...962L..40C,Alonso-Alvarez:2024gdz})
and solitonic cores~\cite{Aghaie:2023lan} (see App.~\ref{app:soliton}) that rely on assumptions about DM abundance.
Detecting or ruling out the predicted turnover with upcoming precision PTA observations would thus isolate
the dynamical imprint of superradiant clouds and pin down ULB mass scales.

Our results connect particle physics, astrophysics, GW astronomy  and cosmology. In particle physics, the derived ULB mass bounds inform future searches beyond the DM paradigm. In astrophysics, boson cloud drag accelerates SMBH coalescence linking to the final-parsec evolution problem. In GW astronomy and cosmology, the predicted GW background suppression and turnover is directly testable with upcoming PTA data.

\section{Conclusions}
We have introduced a novel DM–independent dynamical probe of ULBs based on orbital dynamics. Applying it to a century-monitored  OJ287 system, no excess over relativity predictions excludes ULBs in the mass range
$ (8.5\text{-}22)\times10^{-22}\,\mathrm{eV}$ setting the first DM-independent ULB dynamical constraint derived from binary BH dynamics rather than spin statistics.  
Our framework enables accelerating supermassive‑BH binaries through the long standing final‑parsec evolution bottleneck.  
Consequent suppression and turnover of the nanohertz gravitational wave background provide a decisive test for forthcoming PTA and high‑precision orbital monitors.  
Our framework thus links distinct fields of particle physics, BH astrophysics, and gravitational wave cosmology in a single observational channel.

\medskip\noindent\textit{Acknowledgments.---} 
We thank Masha Baryakhtar and Yong Tang for comments. H.Y.Z. would like to thank Sida Lu and Yi Wang for the fruitful discussion on final-parsec problem. Q.D. and H.Y.Z. were supported by IBS under the project code, IBS-R018-D3. M.H. was supported by IBS under the project code, IBS-R018-D1. V.T. acknowledges support by the World Premier International Research Center Initiative (WPI), MEXT, Japan and JSPS KAKENHI grant No. 23K13109.

\appendix

\section{Scalar-Field Superradiant Gravitational Atom}\label{Sec.GA_Detail}  

A scalar field of mass $\mu$ in the Kerr background of a rotating BH of
mass $M_{\rm BH}$ and dimensionless spin $\chi$
obeys the Klein-Gordon equation
\begin{equation} \label{eq:KG}
\Big(\Box_{\rm Kerr}-\mu^{2}\Big) \Phi=0~, 
\end{equation}
where $\Box_{\rm Kerr}$ is the d'Alembertian operator associated with the Kerr metric. With appropriate boundary conditions,
Eq.~\eqref{eq:KG} separates into Teukolsky radial and angular
equations describing perturbations of a Kerr BH. This can be solved numerically with Leaver’s method
~\cite{Teukolsky:1973ha,Leaver:1985ax}. We consider analytical solutions in the
non-relativistic limit.

For the gravitational fine-structure constant
$\alpha=\mu M_{\rm BH}\ll1$ one can factor out the fast oscillating phase
\begin{equation}
\Phi(t,\mathbf x)=\frac{e^{-i\mu t}}{\sqrt{2\mu}}\,
\psi(t,\mathbf x)+\mathrm{c.c.} ~,
\end{equation}
such that Eq.~\eqref{eq:KG} reduces at leading order in $\alpha$ to the
hydrogen atom analogue Schrödinger-like equation describing the ``gravitational atom (GA)''
\begin{equation}
i\partial_t\psi=\left(-\frac{\nabla^{2}}{2\mu}-\frac{\alpha}{r}
+{\cal O}(\alpha^{2})\right)\psi ~.
\label{eq:Schrodinger}
\end{equation}
Then, the bound GA states are labeled by quantum numbers $(n,\ell,m)$  with wavefunctions
\begin{equation}
\psi_{n\ell m}=R_{n\ell}(r)\,Y_{\ell m}(\theta,\phi)\,
e^{-i(\omega_{n\ell m}-\mu)t}~,
\end{equation}
where $R_{n\ell}$ denotes radial functions and
$Y_{\ell m}$ spherical harmonics.

The GA eigenfunctions are described by complex eigenfrequencies $\omega_{n\ell m}=E_{n\ell m}+i\Gamma_{n\ell
m}$. 
With Detweiler approximation~\cite{Detweiler:1980uk}, considering $ \alpha \lesssim 0.3 $~\cite{Brito2015b,Cannizzaro:2023jle}, the real and imaginary components are analytically expressed as  
\begin{widetext}
\begin{align}\label{eq:wcomp}
\begin{cases}
E_{n\ell m} =\mu\Bigl[1-\dfrac{\alpha^{2}}{2n^{2}}
-\dfrac{\alpha^{4}}{8n^{4}}
-\dfrac{(3n-\ell-1)\alpha^{4}}{n^{4}(\ell+1/2)} +\dfrac{2\chi m\alpha^{5}}{n^{3}\ell(\ell+1/2)(\ell+1)}
+{\cal O}(\alpha^{6})\Bigr],\\ 
\Gamma_{n\ell m} = 2\,\tilde r_{+}\,C_{n\ell}\,g_{\ell m}\bigl(m\Omega_{\rm BH}-\omega_{n\ell
m}\bigr)\,\alpha^{4\ell+5},
\end{cases}
\end{align}
\end{widetext}
where $\tilde r_{+}$ denotes outer horizon of the BH, $C_{nl}$ are numerical coefficients, and  $ g_{lm} = g_{lm}(\chi,\alpha,\mu) $ are functions defined in e.g.~\cite{Detweiler:1980uk,Baumann:2018vus,Baumann:2019eav}.
Exponential superradiant growth occurs when condition $m\Omega_{\rm BH}>\omega_{n\ell m}$ is
satisfied, which happens first for the fastest-growing mode $|211\rangle$. Higher GA
$\ell$-modes grow significantly more slowly with $\Gamma\propto\alpha^{4\ell+5}$ since $\alpha \ll 1$. Hence,
the BH retains finite spin over cosmic times.

The cloud itself can slowly deplete via GW emission at a rate 
\begin{equation}
    \Gamma_{\rm GW}(n,l,m)=-B_{nl}\frac{S_C/m}{M^2}\mu\alpha^{4l+10} ~ ,  
\end{equation}
where $S_C$ is the cloud angular momentum and numerical coefficient $B_{nl}$ can be found in \cite{yoshino2014gravitational}
. For the fastest growing $|211\rangle$ the emission rate is around $\Gamma_{\rm GW}\simeq -2.11 \times 10^{-22}\,{\rm s}^{-1}$. Hence, the respective depletion time exceeds the age of the universe $\tau_U$.  

The cloud density $\rho_{nlm}=M_{nlm} |\psi_{nlm}|^{2}$ can be written as
\begin{equation}
\rho_{nlm}(r,\theta)=\beta\,A_{nlm}(x,\theta) \dfrac{M_{\rm BH}}{r_0^{3}} 
\end{equation} 
with
$x= r/r_0$, $r_0=(\mu\alpha)^{-1}$ being the Bohr radius, and
$\beta= M_{nlm}/M_{\rm BH}$.

\section{Impact of Binary Companion on Gravitational Atoms}\label{Sec.Binary_Companion}
In BH binaries, boson cloud states  could be affected by a companion BH.

A companion of mass $M_\ast$ at an orbital separation $R_\ast(t)$
can perturb the boson cloud gravitationally.
In the reference frame of central BH, considering  $\vec{r}=\{r,\theta,\phi\}$ and keeping only the
Newtonian contributions of the companion’s potential, the gravitational perturbation due to binary companion is
\begin{align}\nonumber\label{eq:Vstar}
V_\ast(t,\vec{r})
=-\alpha\,q\sum_{\ell_\ast\ge2}\sum_{|m_\ast|\le \ell_\ast}&
\mathcal E_{\ell_\ast m_\ast}(\iota_*,\varphi_*)\,
Y_{\ell_\ast m_\ast}(\theta,\phi)\\
&\times F_{\ell_\ast}(r)~,
\end{align}
where
\begin{equation}
F_{\ell_\ast}(r)=
\frac{r^{\ell_\ast}}{R_\ast^{\ell_\ast+1}}\,
\Theta(R_\ast-r)+
\frac{R_\ast^{\ell_\ast}}{r^{\ell_\ast+1}}\,
\Theta(r-R_\ast) 
\end{equation}
and $\Theta$ is the Heaviside step-function, $q= M_{\ast}/M_\text{BH}$ is the mass ratio of the companion and the GA. 
Here we have also introduced the tidal moments 
\begin{equation}
	\mathcal{E}_{l_{\ast}m_{\ast}}=\frac{4\pi}{2l_{\ast}+1}Y^{\ast}_{l_{\ast}m_{\ast}}(\theta_{\ast},\phi_{\ast}) ~ . 
\end{equation}
The angle $\theta_{\ast}$ represents the angle between the BH spin direction and the companion location, while $\phi_{\ast}$ represents the projection of the angular phase of the companion orbit with respect to the primary BH
onto the BH equatorial plane. In the Fermi normal reference frame the dipole
contribution ($\ell_\ast=1$) vanishes
~\cite{Baumann:2018vus,Baumann:2019ztm}, although~\cite{Tomaselli:2023ysb} claims that a residual dipole might exist. Here, we focus on contributions
starting with the quadrupole ($\ell_\ast=2$).

The companion perturbation can induce mixing of different GA eigenstates. Considering boson cloud states
$|n\ell m\rangle$ and $|n'\ell' m'\rangle$
the matrix element factorizes,
\begin{align}
\langle n'\ell' m'|V_\ast|n\ell m\rangle
&=(-1)^{m'+1}\alpha q
\sum_{\ell_\ast,m_\ast}
\mathcal E_{\ell_\ast m_\ast}\,
\mathcal G_{-m'm_\ast m}^{\ell'\ell_\ast\ell}\,I_r ,
\label{eq:matrixelement}
\end{align}
where $I_r$ represents the radial integral
\begin{align}\nonumber
	\nonumber I_r&=\,\int_0^{R^{\ast}} r^2{\rm{d}}rR^{\ast}_{n'l'}(r)R_{nl}(r)\frac{r^{l_{\ast}}}{R_{\ast}^{l_{\ast}+1}}\\
    &+\int_{R^{\ast}}^\infty r^2{\rm{d}}rR^{\ast}_{n'l'}(r)R_{nl}(r)\frac{R_{\ast}^{l_{\ast}}}{r^{l_{\ast}+1}} ~ .
\end{align}
Here, $\mathcal{G}^{l'l_{\ast}l}_{-m'm_{\ast}m}$ is the Gaunt integral
\begin{equation}
	\mathcal{G}^{l'l_{\ast}l}_{-m'm_{\ast}m}=\int{\rm{d}}\Omega Y_{l'-m'}(\Omega)Y_{l_{\ast}m_{\ast}}(\Omega)Y_{lm}(\Omega) ~ . 
\end{equation}
Requiring that Gaunt integral does not vanish imposes that the following selection rules hold
$-m'+m_\ast+m=0$,
$\lvert\ell-\ell'\rvert\le\ell_\ast\le\ell+\ell'$, and
$\ell+\ell_\ast+\ell' =2p, ~p\in \mathbb{Z}$.

With the Newtonian potential, we can construct the new Hamiltonian $H=H_0+V_{\ast}$ consisting of a free part 
\begin{equation}
    H_0= -\frac{1}{2\mu}\partial^2_{\mathbf{r}}-\frac{\alpha}{r}+\mathcal{O}(\alpha^2) ~ , 
\end{equation}
and a tidal perturbation part $V_{\ast}$. The existence of $V_{\ast}$ leads to three phenomena: direct off-resonance coupling of eigenmodes, Landau-Zener (LZ) resonance transitions, and ionization. 

For illustration, we consider two cloud states  $|1\rangle$ and $|2\rangle$ mixed by the
tidal potential $V_\ast$ of the companion. We define the state coupling as  $\eta(t)= V_{21}\langle 2|V_\ast|1\rangle$
and take $m_{1,2}$ to be the azimuthal quantum numbers of the two levels. 
The Hamiltonian in the interaction basis is
\begin{align}\nonumber
H=&
\begin{pmatrix}
\omega_1+V_{11} & V_{12}\\[2pt]
V_{21}          & \omega_2+V_{22}
\end{pmatrix}
\\
=&
\begin{pmatrix}
\bar E_1(t)+i\Gamma_1 & \eta^{*}(t)\\[2pt]
\eta(t)               & \bar E_2(t)+i\Gamma_2
\end{pmatrix},
\label{eq:H_twolevel}
\end{align}
where $\bar E_{1,2}$ and $\Gamma_{1,2}$ are the
unperturbed energy and growth rates, respectively.

\begin{figure*}[t] \centering
        \includegraphics[width=18cm]{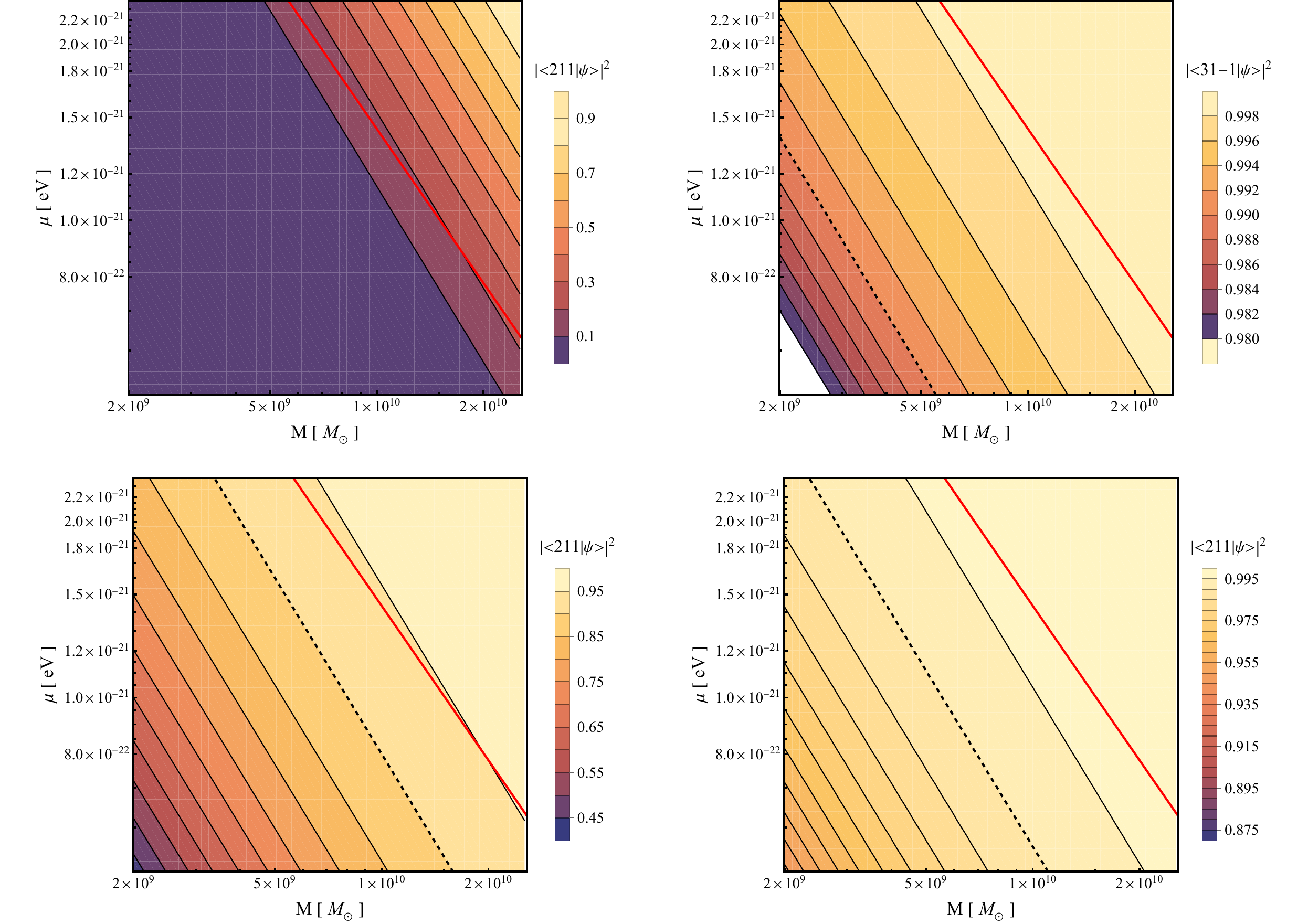}
	\caption{
Survival probability of the GA superradiant eigenmodes under successive resonances.  The red curve marks $170 \times \Gamma_{211}^{-1}=\tau_{\rm U}$, above which the $|211\rangle$ cloud can form within a Hubble time.  [Top Left] $|211\rangle\to|31-1\rangle$ transition at $q=1$. [Top Right] $|31-1\rangle\to|62-2\rangle$ at $q=1$, with the black dashed line showing the 99\% survival contour. [Bottom Left] $|211\rangle\to|31-1\rangle$ at $q=0.1$. The dashed curve is the 90\% contour. [Bottom Right] $|211\rangle\to|31-1\rangle$  with the mass ratio of OJ287 binary, with the 99\% survival contour.
	}
        \label{fig:SurvRate}
\end{figure*}

Transforming to the frame co-rotating with the companion gives the leading modification of the superradiant
rate of the growing level~\cite{Tong:2022bbl,Zhu:2024bqs}. Taking $|1\rangle$ to represent the superradiant state such as $|211\rangle$  and $|2\rangle$ to represent the absorptive state such as $|21-1\rangle$, the instantaneous correction to superradiant rate of state $|1\rangle$ is  
\begin{equation} \label{eq:DGamma}
\Delta\Gamma_1\simeq
-\,
\frac{\Gamma_1-\Gamma_2}
     {\bigl[(E_1-E_2)-(m_1-m_2)\dot\varphi_\ast(t)\bigr]^2}
\,|\eta(t)| ,
\end{equation}
which is always negative and largest near
resonance.  For a general binary orbit and eccentricity, one should average over a period~\cite{Fan:2023jjj}
\begin{equation}
\langle\Delta\Gamma_1\rangle=
\frac{1}{T}\int_{0}^{2\pi}
\Delta\Gamma_1\,
\frac{d\varphi_\ast}{\dot\varphi_\ast},
\label{eq:DGamma_orbavg}
\end{equation}
which shows that the effect scales with $|\eta|$.

In the case of OJ287, the orbital angular momentum is nearly orthogonal to the BH spin, \textit{i.e.} inclination angle $\iota\simeq90^{\circ}$ (see schematic Fig.~\ref{fig:illustration}). This strongly suppresses the overlap
$|\eta|$ and correction to the superradiant rate.  We estimate
$\langle\Delta\Gamma_1\rangle/\Gamma_1\ll 1$. Thus, off-resonant
mixing can be safely neglected in our analysis.

The LZ resonance transitions constitute other prominent effects. We observe that Eq.~\eqref{eq:DGamma}  develops a pole when the driving frequency
$\dot\varphi_\ast$ sweeps through
$\Omega_r=(E_1-E_2)/(m_1-m_2)$.
Close to the crossing we linearize the chirp,
$\dot\varphi_\ast (t) \simeq \dot\varphi_\ast(t_0)+\gamma t$ where $\gamma = \gamma_\text{GW} + \gamma_\text{DF} + \gamma_\text{BR}$ takes into account the energy loss via GW, dynamical friction and back-reaction of resonance effect on the companion. Here $\gamma_\text{GW}$ can be calculated Eq. (2.28) in \cite{Baumann:2019ztm} and $\gamma_\text{DF}$ can be estimated by $\langle P_\text{DF}\rangle/\langle P_\text{GW} \rangle \times \gamma_\text{GW} $. Since the back-reaction effect is subdominant, we ignore the $\gamma_\text{BR}$ in our calculation. 
and treat the two-level system with the standard
LZ formalism
~\cite{Baumann:2019ztm,Tomaselli:2024bdd}.  
With initial populations
$c_1(-\infty)=1$, $c_2(-\infty)=0$ the occupation after the passage is  
\begin{equation}
|c_1|^{2}=e^{-2\pi Z},\quad
|c_2|^{2}=1-e^{-2\pi Z},\quad
Z=\frac{|\eta|^{2}}{\gamma},
\label{eq:LZ}
\end{equation}
where $Z$ is the usual LZ parameter.

For a saturated $|211\rangle$ cloud three LZ resonances are in principle
relevant:
$|211\rangle\to|31{-}1\rangle\to|62{-}2\rangle\to|200\rangle$
~\cite{Baumann:2019ztm}.
The last state is rapidly absorbed, which we estimate as $\Gamma_{200}\sim10^{-11}\,\mathrm{s^{-1}}$ for benchmark $\alpha=0.124$ as relevant for OJ287 primary source. 
Thus, the boson cloud will be depleted in roughly $\sim 3 \times 10^3$ years once it transitions to the $|200\rangle$ state. However, during these resonances, the binary system is situated within the cloud. Full state depletion would require $Z\gg1$ in the preceding evolution steps.
In realistic binaries the external energy loss is dominated by
GW radiation and the cloud’s dynamical friction,
making $\gamma$ large and $Z\lesssim10^{-3}$.
Consequently each resonance is highly non-adiabatic and transfers at
most a few per~cent of the state.
This significantly suppresses the LZ parameter $Z$, resulting in an incomplete resonance transition.

In Fig.~\ref{fig:SurvRate} we display the survival probability of the
$|211\rangle$ and $|31{-}1\rangle$ levels after the chain of Bohr-type
LZ crossings, as a function of $(M_{\rm BH},\mu)$ and for
several characteristic BH binary mass ratios $q$. 
For equal masses ($q=1$) the first
crossing almost completely transfers state population from $|211\rangle$ to
$|31{-}1\rangle$. However, $|31{-}1\rangle$ itself survives the
subsequent crossing at the $>99\%$ level, and the density profiles of
the two states differ by less than an order of magnitude
as shown on Fig.~\ref{fig:DensityProf}.  When the companion is lighter,
e.g.\ $q=0.1$, more than $90\%$ of the cloud remains in
$|211\rangle$ over the entire parameter space, and in the residual
region even a complete transfer to $|31{-}1\rangle$ would leave that
state $100\%$ intact considering $|31-1\rangle\to|62-2\rangle$ resonance.  Considering OJ287 mass ratio
$q\simeq 8 \times 10^{-3}$ the survival of $|211\rangle$ state exceeds $99\%$ throughout its history. Hence,  resonance depletion can
be neglected in the analysis of OJ287.

Fine- and hyperfine-structure transitions require binary separations well
above those reached by the OJ287 binary throughout its evolution history, for example, tracing back to
redshift $z\simeq10$, and therefore never occur.  Back-reaction of the
few-percent state population that is shuffled between levels is negligible
compared with the dominant dynamical-friction torque. Allowing for
non-vanishing eccentricity or a 
$ \iota \simeq 90^\circ $ inclination  
results in even greater survival fraction~\cite{Tomaselli:2023ysb,Tomaselli:2024bdd}.
We therefore consider, for simplicity, a circular coplanar orbit in the main text.

Apart from the bound-bound resonance transition, there is also a resonance transition from a bound state to an unbound state, which is called ionization~\cite{Baumann:2021fkf,Tomaselli:2023ysb}. The back-reaction of ionization was shown to be well described as dynamical friction~\cite{Tomaselli:2023ysb}. Ionization can eject bosons and deplete the cloud, which, however, is subdominant compared with the accretion effect of the companion for a massive cloud such as the case we consider here~\cite{Baumann:2021fkf}. A rough estimate of the timescale for accretion to consume the cloud is $ \gtrsim 10^8 {\rm yr} $, which is much larger than the duration of accretion experienced by the companion so far. We also estimate that tidal effects caused by the companion~\cite{Cardoso:2020hca} result in a small tidal parameter $ \epsilon \sim 10^{-8}$ for OJ287, indicating stability to them.
Hence, dominant depletion channels are insignificant, and GA cloud can be considered as quasi-static, consistent with our analysis. 

\section{SMBH binary merger rate}\label{app:SMBH_population}

We can estimate the SMBH binary merger rate by mapping galaxy mergers to SMBH coalescences, considering each galaxy hosts a central BH.  For primary SMBH mass by $M_{\rm BH}$ and mass ratio by $q$, and the primary galaxy stellar mass by  $M_{\rm g}$ with merging galaxy mass ratio $q_{\rm g}$, one has
\begin{equation}
  \frac{\dd^3n}{\dd z\,\dd q\,\dd M_{\rm BH}}
  = \frac{\dd^3n_{\rm g}}{\dd z\,\dd q_{\rm g}\,\dd M_{\rm g}}
    \,\frac{\dd M_{\rm g}}{\dd M_{\rm BH}}\,\frac{\dd q_{\rm g}}{\dd q}\,.
\end{equation}
The galaxy merger rate is parametrized as~\cite{NANOGrav:2023hfp,Chen:2018znx}
\begin{equation}
  \frac{\dd^3n_{\rm g}}{\dd z\,\dd q_{\rm g}\,\dd M_{\rm g}}
  = \frac{\Psi(M_{\rm g},z')}{M_{\rm g}}\,
    \frac{\mathcal{P}(M_{\rm g},q_{\rm g},z')}{T_{\rm g-g}(M_{\rm g},q_{\rm g},z')}
    \,\frac{\dd t}{\dd z'}\,,
\end{equation}
where $\Psi$ is the galaxy stellar mass function, $\mathcal{P}$ the galaxy pair fraction, and $T_{\rm g-g}$ the galaxy merger timescale.  The cosmic time-redshift relation $\dd t/\dd z$ in a flat $\Lambda$CDM cosmology is given by
\begin{align}
  \frac{\dd t}{\dd z} &= \frac{1}{(1+z)\,H(z)}~,\\
  H(z)&=H_0\sqrt{\Omega_\Lambda + \Omega_m(1+z)^3}~,
\end{align}
with Planck~2018 parameters $H_0=67.4\,$km\,s$^{-1}$\,Mpc$^{-1}$, $\Omega_m=0.315$, $\Omega_\Lambda=0.685$~\cite{Planck:2018vyg}.

Adopting the Schechter form for galaxy stellar mass function, we parametrize the relevant quantities as
\begin{align}
    \Psi(M_{\rm g}, z) &= \Psi_0 \left(\frac{M_{\rm g}}{M_\Psi}\right)^{s_\Psi} \exp \left(- \frac{M_{\rm g}}{M_\Psi}\right)~, \\
    \mathcal{P}(M_{\rm g}, q_{\rm g}, z) &= \mathcal{P}_0 (1+z)^{\beta_{p_0}}~, \\
    T_\mathrm{g-g}(M_{\rm g}, q_{\rm g}, z) &= T_0 (1 + z)^{\beta_{t_0}} q_{\rm g}^{\gamma_{t_0}}~,
\end{align}
with
\begin{align}
    \log_{10}\left(\frac{\Psi_0}{\mathrm{Mpc}^{-3}}\right) &= \psi_0 + \psi_z z~, \\
    \log_{10}\left(\frac{M_\Psi}{M_\odot}\right) &= m_{\psi_0} + m_{\psi_z} z~, \\
    s_\Psi &= 1 + s_{\psi_0} + s_{\psi_z} z~.
\end{align}
We can further relate $M_{\rm g}$ to the SMBH mass via the $M_{\rm BH}$--$M_{\rm bulge}$ relation~\cite{Kormendy:2013dxa}
\begin{align}\label{eq:BH_bulge}
    \log_{10}\left(\frac{M_{\rm BH}}{M_\odot}\right) = 8.7 + 1.1 \log_{10}\left(\frac{M_\mathrm{bulge}}{10^{11} M_\odot}\right)~,
\end{align}
where the bulge mass can be expressed by the stellar mass of galaxy as~\cite{Lang:2014cta, Bluck:2014ila}
\begin{align}\label{eq:bulge_star}
    M_\mathrm{bulge} = f_{\rm g} M_{\rm g}~.
\end{align}
From Eqs.~\eqref{eq:BH_bulge} and \eqref{eq:bulge_star}, the relation between the mass ratios $q_{\rm g}$ and $q$ can be found as 
\begin{align}
    q = q_{\rm g}^{1.1}~.
\end{align} 

In Tab.~\ref{Table.GW_para} we display the numerical values of relevant parameters~\cite{NANOGrav:2023hfp, Alonso-Alvarez:2024gdz}. 

\begin{table}[t]
\centering
\begin{tabular}{    c | c || c | c    } 
 \hline\hline
 Parameter &   Value  & Parameter &   Value  \\
 \hline
 $\psi_0$   & $-2.63$ & $\mathcal{P}_0$ & $0.033$\\
 $\psi_z$ &  $-0.6$ & $\beta_{p_0}$ & $1$ \\
 $m_{\psi_0}$ & $11.5$ & $T_0$ (Gyr) & $0.5$ \\
 $m_{\psi_z}$ & $0.11$ & $\beta_{t_0}$ & $-0.5$ \\
 $s_{\psi_0}$ & $-1.21$ & $\gamma_{t_0}$ & $-1$\\
 $s_{\psi_z}$ & $-0.03$ & $f_{\rm g}$ & $0.615$\\
 \hline\hline
\end{tabular}
 \caption{Parameter values used for evaluating the
 the SMBH binary merger rate.}
 \label{Table.GW_para}
\end{table}

\section{Statistical Analysis of GW Background Fit}\label{app:chi_squared}
To evaluate the performance of stochastic gravitational wave background in superradiant boson cloud scenario, we estimate the value of $\chi^2$ in both superradiant boson cloud scenario and pure GW scenario. The $\chi^2$ can be calculated as
\begin{align}
    \chi^2 = \sum_{i=1}^N \frac{(O_i-E_i)^2}{\sigma_i^2}~,
\end{align}
where $O_i$ is the $i$th observed characteristic strain at measured frequency, $E_i$ is the corresponding expected value of characteristic strain in difference scenarios, and $\sigma_i^2$ is the variance of the $i$th distribution of characteristic strain. $N$ is the total number of observed data points. Give the characteristic strain of PTA data as shown in Table.~\ref{Table.PTA.data}, we have the observed value of characteristic strain $O_i$ and corresponding variance $\sigma_i^2$. Since the value of upper variance $\sigma_{+}^2$ and lower variance $\sigma_{-}^2$ are different, we choose the mean value of upper and lower variance in $\chi^2$ calculation as
\begin{align}
    \sigma^2 = \frac{\sigma_{+}^2 + \sigma_{-}^2}{2}~.
\end{align}
Then we apply Eq.~\eqref{eq:SGWB_strain} to calculate $E_i$ at measured frequency in $O_i$. 
It gives a $\chi^2$ fit result from Fig.~\ref{fig:SGWB} that predicted strain in GA cloud scenario gives $\chi^2 = 16.5$ (blue solid  - all SMBHs with maximal spin), $\chi^2 = 17.5$ (blue dashed  - conservative superradiance timescales), while predicted strain in GW scenario gives $\chi^2 = 18.8$ (black solid -  general relativity prediction).

\begin{table}[t]
\centering
\begin{tabular}{    c | c | c || c | c | c     } 
 \hline\hline
 &&&&&\\[-1em]
 $f \, [\text{yr}^{-1}]$ &   $h_c/10^{-15}$  & Ref & $f \, [\text{yr}^{-1}]$ &   $h_c/10^{-15}$  & Ref\\
 \hline
 &&&&&\\[-1em]
 $0.062$   & $6.5^{+4.5}_{-2.4}$ & \multirow{5}{*}{NANOGrav}&$0.055$ & $8.3^{+6.8}_{-3.4}$ & \multirow{11}{*}{PPTA}\\
 &&&&&\\[-1em]
 $0.12$ &  $7.9^{+3.0}_{-1.8}$ & & $0.11$ & $9.6^{+4.2}_{-3.6}$ \\
 &&&&&\\[-1em]
 $0.19$ & $7.4^{+3.1}_{-2.0}$ & & $0.17$ & $7.4^{+3.1}_{-1.8}$ \\
 &&&&&\\[-1em]
 $0.25$ & $6.4^{+3.3}_{-1.9}$ & & $0.22$ & $6.3^{+4.5}_{-2.5}$ \\
 &&&&&\\[-1em]
 $0.31$ & $9.3^{+4.8}_{-4.2}$ & & $0.28$ & $1.0^{+3.2}_{-0.8}$\\
 &&&&&\\[-1em]
 \cline{1-3}
 &&&&&\\[-1em]
 $0.097$   & $8.0^{+4.0}_{-2.7}$ & \multirow{6}{*}{EPTA}& $0.33$ & $6.3^{+4.4}_{-4.0}$\\
 &&&&&\\[-1em]
 $0.19$ &  $9.6^{+2.9}_{-1.9}$& & $0.39$ & $3.7^{+3.7}_{-1.5}$ \\
 &&&&&\\[-1em]
 $0.29$ & $8.2^{+3.8}_{-8.2}$ & & $0.44$ & $7.1^{+3.9}_{-2.6}$ \\
 &&&&&\\[-1em]
 $0.39$ & $11.0^{+4.5}_{-2.9}$ & & $0.50$ & $1.6^{+3.2}_{-1.0}$ \\
 &&&&&\\[-1em]
 $0.48$ & $5.2^{+10.0}_{-5.5}$ & & $0.55$ & $3.6^{+3.6}_{-2.09}$\\
 &&&&&\\[-1em]
 $0.58$ & $0.6^{+4.8}_{-0.4}$ & &  &  \\
 \hline\hline
\end{tabular}
 \caption{The data points of the PTA characteristic strain in NANOGrav \cite{NANOGrav:2023gor}, EPTA \cite{EPTA:2023xxk} and PPTA \cite{Reardon:2023gzh}.}
 \label{Table.PTA.data}
\end{table}

\section{Boson Solitons and Gravitational Waves}\label{app:soliton}

\begin{figure}[htpb] \centering
        \includegraphics[width=8cm]{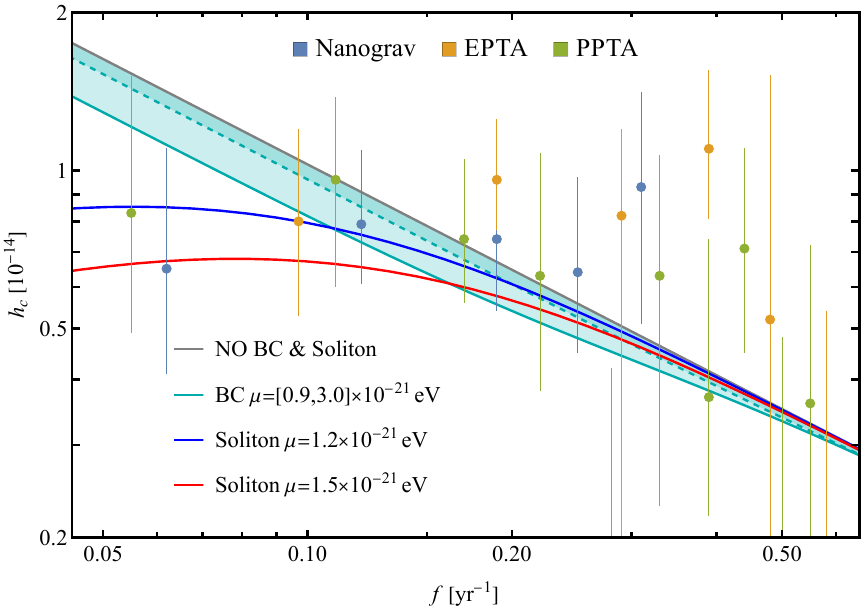}
	\caption{
	The characteristic strain of GW background produced via SMBH binaries with different impacts from ULB soliton cores and superradiant GA boson clouds. We consider GA boson clouds in $|211\rangle$ state, with critical SMBH mass satisfying $\Gamma_{211} = 1/ \tau_\mathrm{U}$ for green solid line and $\Gamma_{211} = 170/ \tau_\mathrm{U}$ for green dashed line.    
    The shaded region depicts impact from superradiant boson clouds on SGWB, considering ULB masses in the range of $(0.9-3) \times 10^{-21}\,\text{eV}$. Impact of solitonic cores with ULB masses of $\mu = 1.2, 1.5 \times 10^{-21} \, \mathrm{eV}$ is shown. The mass ratio $\varepsilon$ between soliton core and SMBH is set as $10^{-3}$. The observation data points come from NANOGrav~\cite{NANOGrav:2023gor}, EPTA~\cite{EPTA:2023xxk}, and PPTA~\cite{Reardon:2023gzh} are shown.
	}
        \label{fig:SGWB_soliton}
\end{figure}

ULB DM can form compact solitonic cores around SMBHs. The resulting signatures imprinted on stochastic GW background allows restricting ULB masses in the range of $1.3 \times 10^{-21} \, \mathrm{eV}$ to $1.4 \times 10^{-20} \, \mathrm{eV}$~\cite{Aghaie:2023lan}. We calculate
dynamical friction $\langle P_\mathrm{DF}\rangle$  considering density profile of solitons described as~\cite{Aghaie:2023lan}  
\begin{align}
    \rho_\mathrm{sol} \simeq \rho_0 e^{-2 \alpha\mu r}
\end{align}
with
\begin{align}
    \rho_0 \simeq 5 \times 10^4 \varepsilon \left(\frac{M}{10^8 \, M_\odot}\right)^4 \left(\frac{\mu}{10^{-21} \, \mathrm{eV}}\right)^6 \, M_\odot \, \mathrm{pc}^{-3}~,
\end{align}
where $\varepsilon$ is the mass ratio between solitonic core and SMBH.

In Fig.~\ref{fig:SGWB_soliton} we display the GW strain for stochastic background from SMBH solitonic cores for various ULB DM masses considering population of SMBHs following Eq.~\eqref{eq:SGWB_strain} and SMBH mass distribution in the range of $10^5 - 10^{10} M_{\odot}$. We compare these results with dynamical friction effects from ULB GA boson clouds in $|211\rangle$ state considering SMBHs with mass greater than critical mass $M_\mathrm{crit}$ that satisfies condition $\Gamma_{211} = 1/ \tau_\mathrm{U}$.
The dynamical friction effects from ULB‐DM solitonic cores and from superradiant clouds imprint qualitatively different low‐frequency turnovers in the PTA spectrum, allowing these two scenarios to be distinguished.

\bibliographystyle{utphys}
\bibliography{references}

\end{document}